# A 0.042 mm² programmable biphasic stimulator for cochlear implants suitable for a large number of channels

W. Ngamkham[1], M. N. van Dongen[1], W. A. Serdijn[1], C. J. Bes[1], J. J. Briaire[2], J. H. M. Frijns[2]

[1]*Section Bioelectronics, Faculty of Electrical Engineering, Mathematics and Computer Science, Delft University of Technology*
*Mekelweg 4, 2628CD Delft, The Netherlands*
Email: see http://bioelectronics.tudelft.nl

[2]*Leiden University Medical Centre (LUMC), P.O. Box 9600, 2300 RC Leiden, The Netherlands*

**Abstract**

This paper presents a compact programmable biphasic stimulator for cochlear implants. By employing double-loop negative feedback, the output impedance of the current generator is increased, while maximizing the voltage compliance of the output transistor. To make the stimulator circuit compact, the stimulation current is set by scaling a reference current using a two stage binary-weighted transistor DAC (comprising a 3 bit high-voltage transistor DAC and a 4 bit low-voltage transistor DAC). With this structure the power consumption and the area of the circuit can be minimized. The proposed circuit has been implemented in AMS 0.18$\mu$m high-voltage CMOS IC technology, using an active chip area of about 0.042mm². Measurement results show that proper charge balance of the anodic and cathodic stimulation phases is achieved and a dc blocking capacitor can be omitted. The resulting reduction in the required area makes the proposed system suitable for a large number of channels.

## 1. Introduction

A cochlear implant (CI) is a surgically implanted electronic device that bypasses the damaged parts of the inner ear and directly stimulates the remaining hearing nerve fibers in the cochlea with electrical signals [1]. In general, a cochlear implant consists of an external part and an internal part. The external part comprises: 1) a microphone, which picks up sound from the environment; 2) a speech processor which selectively filters sound to prioritize audible speech, splits the sound into channels and sends the electrical sound signals through a thin cable to the transmitter; and 3) a transmitter, which is a coil held in position by a magnet placed behind the outer ear, that transmits power and the processed sound signals through the skin to the internal part. The internal part comprises: 1) a receiver and a stimulator that convert the signals into electric impulses and sends them through an internal cable to the electrodes; 2) an electrode array inserted inside the cochlea, which sends the impulses to the nerves in the scala tympani and then directly to the brain through the auditory nerve system. Generally, users can understand speech by using the device in a clean environment, i.e., without background noise or when communicating via the telephone.

This paper will focus on the stimulator that is responsible for generating electric pulses. Since the stimulator is implanted inside the body, a small size and a low power consumption are critical requirements, especially if a large number of channels is preferred. Moreover, the circuit must be able to provide charge-balanced stimulation in order to prevent tissue damage [2]. A current mode stimulator seems to be an attractive method because the amount of charge injected into the tissue can easily be defined by the current amplitude and the duration of the pulses.

Several current mode stimulators have been reported thus far. Stimulators based on a current mirror circuit have been widely used [3]-[12]. To maintain constant current stimulation, wide-swing and regulated cascode current mirror topologies are used but these limit the voltage compliance. Moreover, when using a dual supply with two current sources to create the stimulator, additional circuitry to match the two current sources is needed to ensure charge cancellation [4]-[6]. A voltage-controlled resistor based implementation has been presented in [7] to achieve a high voltage compliance but it needs additional circuitry to reduce non-linearity. A blocking capacitor free stimulator using dynamic current matching is a useful idea to reduce the size of the implant and preserve a charge error less than 6pC [8]. However, a dual supply is used and additional circuitry is needed in this method.

Another issue is related to making the output current programmable. In many cases a digital-to-analog converter (DAC) is used to generate a programmable reference current [3]-[5], [9],[10]. Then, a current mirror replicating or scaling the reference current is used to provide the stimulation current. All these system blocks consume power and area. It has therefore been suggested to combine the DAC function into the output current stage in order to reduce the complexity and minimize the silicon area and power consumption [11]. Our work has adopted this suggestion and tackles several other drawbacks mentioned in the previous paragraph.

In this paper, a current mode, biphasic neural stimulator for application in cochlear implants is proposed. It uses a compact stimulator circuit, avoids the use of external blocking capacitors by achieving a good charge balance and thereby allows an increase in the number of stimulator channels. By using a double-loop negative feedback topology, the output impedance of the current source can be maximized while only one effective drain-source voltage drop ($V_{eff}$) is required. This means that more voltage headroom at the tissue is achieved and more charge can be conveyed into the tissue.

The circuit provides a programmable biphasic stimulator that includes the DAC function embedded in the output current stage. By using a two stage binary-weighted transistor DAC configuration (employing high-voltage and low-voltage DAC topologies), the settling time of the stimulation pulse improves, and the area of the circuit is minimized.

Moreover, by using a single current source from a single supply, we achieve charge-balanced stimulation during every stimulation cycle and avoid the need for a dual supply. A switch array is used to reverse the current direction.

The remaining sections of the paper are organized as follows. Section 2 presents the system design of the proposed stimulator. Section 3 describes the circuit implementation and the fabrication of the proposed stimulator. The measurement results are presented in Section 4. The conclusions are given in the final section.

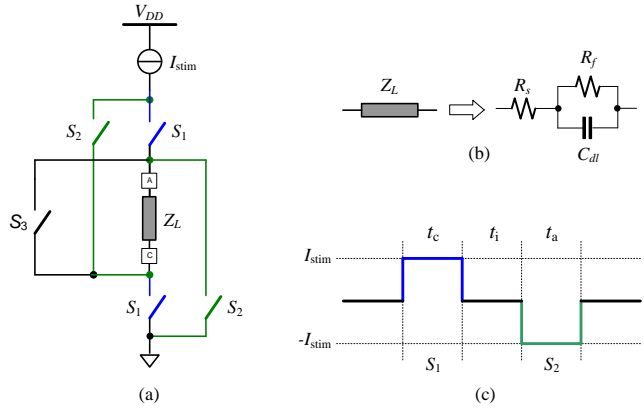

Figure 1: (a) Single supply stimulation scheme. (b) Electrode-tissue interface model. (c) Biphasic stimulation waveform.

## 2. System design

The stimulator uses a single supply and a single current source stimulation scheme as depicted in Fig. 1(a). The configuration consists of three main parts, the load, $Z_L$, representing the tissue and the electrode-tissue interface, the switch array ($S_1$, $S_2$, and $S_3$) and the current source, $I_{stim}$. The individual parts will be described in the following subsections.

### 2.1 Electrode-Tissue Interface Model

The load of the stimulator comprises the tissue and the electrode-tissue interface which can be modeled in the electrical domain. In this work a simple model is used as shown in Fig. 1(b) [13]. $R_s$ corresponds to the resistance of the electrode leads and the tissue. In practice the resistance of the electrolyte (the tissue) will be much higher and is therefore the dominant component. $C_{dl}$ and $R_f$ correspond to the interface between the electrode and the tissue. $C_{dl}$ is the double-layer capacitance and $R_f$ is the Faradaic resistance. For simplify, we assume that $R_f$ remains large (>1MΩ) during the stimulation pulse (<100µs), and can therefore be neglected. The stimulator chip is designed for electrode-tissue impedances in the range of $R_S$=1kΩ~10kΩ, $C_L$=1nF~10nF (RC series impedance in cochlear implants) [8].

### 2.2 Switch Array

The stimulation pattern is controlled by the switch array comprising switches $S_1$, $S_2$, and $S_3$. The principle of constant current biphasic stimulation can be described using Fig. 1(c). When switches $S_1$ are closed ($t_c$) the current flows from A to C. When switches $S_2$ are closed and $S_1$ are opened ($t_a$) the current reverses its direction. An inter-phase delay ($t_i$) is added between the stimulation phases when switches $S_1$ and $S_2$ are opened. Switch $S_3$ is used to short circuit and thus passively discharge the tissue. The advantage of using the switch array for performing both anodic and cathodic current injection is that only a single voltage supply is needed. Also, since only one source is used, the currents are easily matched during both phases.

### 2.3 High Output Resistance Current Source

A straightforward implementation of a current driver uses a current mirror which is often cascoded in order to have a sufficiently high output resistance [14]. Fig. 2(a) shows the simplest version of a PMOS cascode current mirror. A small reference current, $I_{DAC}$, generated from a digital to analog converter, is applied through diode connected transistors $M_1$ and $M_4$. Subsequently, the current is scaled up by a factor $n$ to become the stimulation current, $I_{stim}$, flowing through transistors $M_2$, $M_3$, and load $Z_L$. In this case, the output resistance equals $g_{m3}r_{o3}r_{o2}$, where $r_{o2}$ and $r_{o3}$ are the output resistance of $M_2$ and $M_3$, respectively, $g_{m3}=(2I_{D3}K_pW/L)^{0.5}$ is the transconductance of $M_3$, where $I_{D3}$ is the drain current through $M_3$, $K_p$ is the intrinsic transconductance, and $W$ and $L$ are the width and length of $M_3$, respectively. However, the minimum required voltage across the current source (voltage drop, $V_d$) becomes one source-gate voltage ($V_{SG}$) plus one effective source-drain voltage ($V_{eff}$). This limits the voltage headroom $V_L$ and the amount of charge that can be conveyed to the tissue (load).

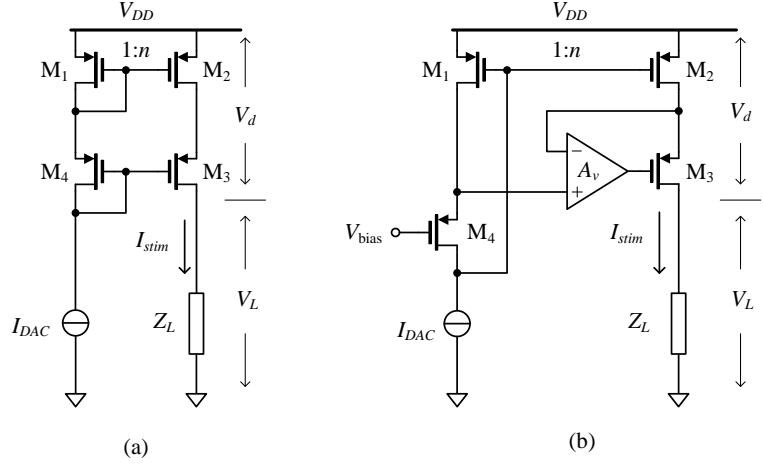

Figure 2: Cascode current sources: (a) simple cascade, and (b) regulated cascade.

To increase $V_L$, a current mirror employing active feedback to boost the output impedance can be used. In Fig. 2(b), a high-gain amplifier, $A_v$, is applied to make the drain voltage of $M_2$ equal to the drain voltage of $M_1$. The same biasing condition makes $I_{stim}$ n times $I_{DAC}$. The output resistance of the active feedback current generator is given by $A_v g_{m3}\ r_{o3}r_{o2}$. This causes $V_d$ to become $V_{eff3} + V_{eff2}$. Due to the fact that the output resistance of this mirror is higher than in the previous cases but requires less $V_d$, this cascoded structure is popularly used in neural stimulation [4],[15].

## 2.4 Proposed Current Source

In order to allow for an even higher $V_L$, we proposed a high output impedance current source that requires $V_d$ to be only a single $V_{eff}$ [16],[17]. The concept of the proposed current source is shown in Fig. 3(a). It contains two feedback loops. The first local one is used for high precision down scaling of $I_{stim}$ (to $I_{stim}/n$). Transistor $M_2$ generates $I_{stim}$ flowing through $Z_L$. The gate terminal of transistor $M_1$ is connected to the gate terminal of $M_2$ to accurately scale down the current flowing through $M_3$ to summing node A. Based on the same principle as used for the current mirror of Fig. 2(b), $A_v$ is used to force the drain terminals of $M_1$ and $M_2$ to be equal, resulting in a very precisely copied current $I_{stim}/n$ flowing into node A. This current will be compared with $I_{DAC}$ and an error current, $I_e$, equal to

$$I_e = \frac{I_{stim}}{n} - I_{DAC}, \qquad (1)$$

flows into transimpedance amplifier $Z_m$ converting the very small $I_e$ into the voltage needed at the gate of $M_2$ to produce $I_{stim}$. $I_e$ will be forced to be zero by the large loop gain of the second global feedback loop creating a relationship of

$$I_{stim} = nI_{DAC}. \qquad (2)$$

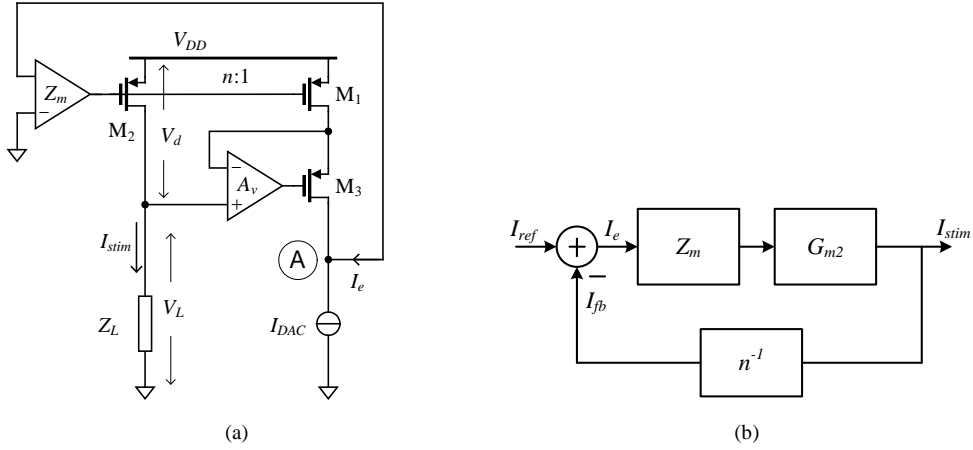

Figure 3: Proposed current source (a) and feedback block diagram of the proposed current source (b).

We can simplify the feedback block diagram of the proposed circuit as shown in Fig. 3(b). The loop gain of the system can be found to be

$$LG = \frac{G_{m2}Z_m}{n}, \qquad (3)$$

where $G_{m2}$ is the transconductance gain of transistor $M_2$.

To maintain the desired current given by (2), $LG$ needs to be as large as possible. Since $G_{m2}$ is limited by the values of $I_{stim}$ and the dimension of $M_2$ and $n$ is preferred to be high (10-100) to keep the total power consumption low, a large $Z_m$ becomes the main factor that defines the accuracy of the proposed circuit.

The simulated output current versus the voltage headroom of the three current source designs $V_L$, the simple cascode (Fig. 2(a)), the regulated cascode (Fig. 2(b)), and the proposed circuit (Fig. 3(a)) is shown in Fig. 4(a). AMS 0.18µm high-voltage technology parameters were used for circuit simulations. The simulation is performed using the same transistor dimensions for all three designs. A high voltage supply (>10V) is needed to accommodate the maximum current required (1mA) through the maximum load expected (< 20kΩ). $V_{DD}$ is set at 18V and $I_{DAC}$ is 50µA, yielding $I_{stim}$=1mA with a scaling factor of $n = 20$. Ideal op-amps with a gain $A_v = 200$ are used for the circuits in Figs. 2(b) and Fig. 3(a). $V_L$ was varied from 0 to 18 V with a 0.5V step size by using an ideal voltage source, as we can see the proposed current source achieves a larger voltage headroom than those of the others. This verifies that the proposed current source can inject more charge into the tissue for the same supply voltage.

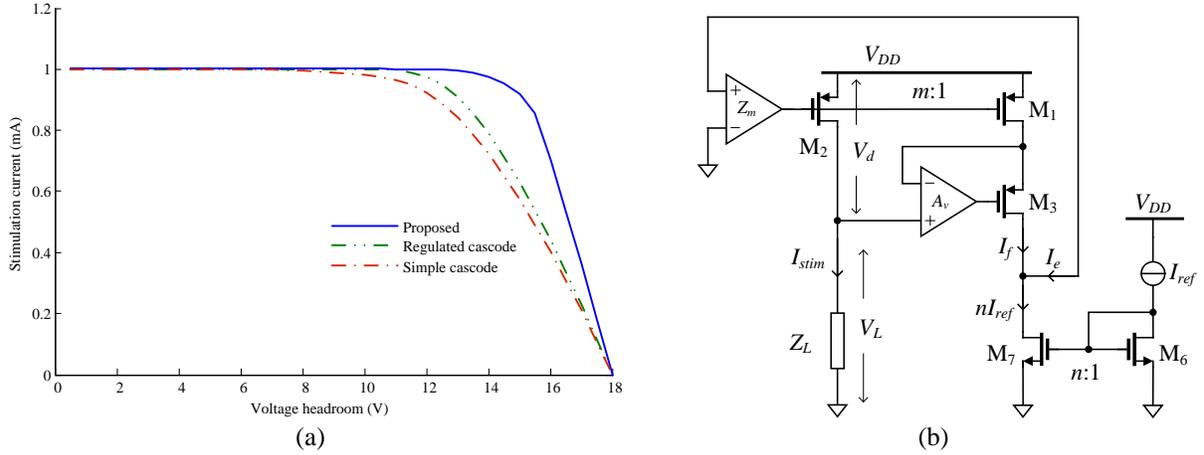

Figure 4: (a) A comparison between the output characteristics of the current sources. (b) Modification of the proposed current source.

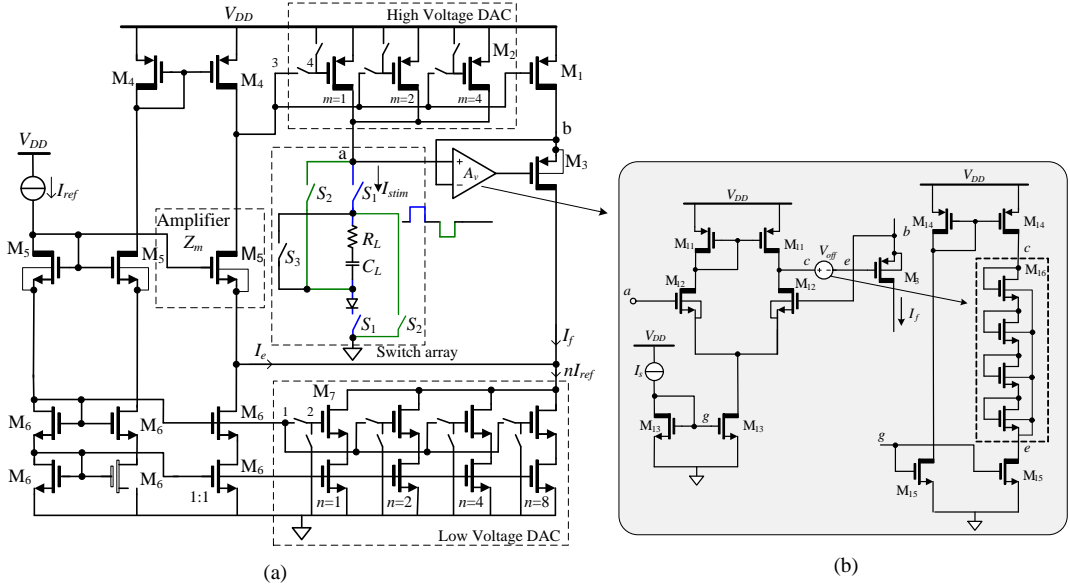

Figure 5: Principle circuit diagram of the implemented programmable biphasic stimulator circuit.

## 3. Circuit implementation

From the concept as described in the previous section, we have derived a programmable biphasic stimulator circuit for cochlear implants. The design aims to support the flexible electrode array as developed in the SMAC-It (Smart cochlear implants) project [18]. In order to reduce the size and parasitic capacitances of the stimulator output transistor ($M_2$), the proposed current source was modified as shown in Fig. 4(b). $I_{stim}$ is created by scaling a reference current ($I_{ref}$) by scaling factors $n$ and $m$ of $M_7$ and $M_2$, respectively. Thereby $I_{stim}$ becomes

$$I_{stim} = n \cdot m \cdot I_{ref}. \qquad (4)$$

To make $n$ and $m$ programmable $M_7$ and $M_2$ are implemented using a binary weighted DAC scheme. The circuit diagram of the implemented stimulator circuit is shown in Fig. 5(a). This requires the use of high-voltage (HV) transistors (indicated by the thick drain terminal) combined with low-voltage (LV) transistors. To minimize the area occupied by the circuit, the number and size of HV transistors used should be as small as possible. The design of the individual sub-circuits will be described in the next subsections.

*3.1 High-Voltage and Low-Voltage DAC Configuration*

In order to create a 10μA resolution for a 1mA full-scale stimulation current, a 7-bit resolution is required. The silicon area of the circuit can be minimized when using two stages in cascade, a HV DAC and a LV DAC. The number of bits in the HV DAC should be as small as possible. This will reduce the number of (large) HV transistors resulting in a smaller area as well as a lower parasitic capacitance. However, a certain minimum equivalent transistor size is needed to be able to supply the maximum stimulation current. In our design 3 bits for the HV DAC was found to be optimal.

The remaining 4 bits can be implemented using LV transistors. These transistors are much smaller, making the area contribution negligible compared to the area occupied by the HV DAC.

The reference current was chosen to be $I_{ref} = 10\mu A$. By enabling one or more transistors in the binary weighted DACs (using transistor switches), $I_{stim}$ can be made programmable using the following relation:

$$I_{stim} = \left(\sum_{u=0}^{2} a_u 2^u\right)\left(\sum_{l=0}^{3} a_l 2^l\right) I_{ref}, \tag{5}$$

in which $u$ and $l$ are the bit-numbers of the enabled HV transistors $M_2$ and LV transistors $M_7$, respectively. In this way the LV DAC can generate a current in steps of 10μA from 10μA to 150μA. The HV DAC can scale this current with a factor 1 up to 7, resulting in a maximum stimulation current of 1.05mA.

*3.2 Differential Amplifier $A_v$*

Amplifier $A_v$ is used in the feedback loop to control the drain voltage of $M_1$. It is implemented using a standard differential amplifier (using HV transistors) with an active load as depicted in Fig. 5(b). An offset voltage source, $V_{off}$, is needed at the output of the amplifier to bias the gate of M3 properly. It was implemented using a diode connected LV transistor chain and current sources $I_s = 10\mu A$. The minimum common mode input voltage that the amplifier can handle is about 3V because of the biasing of the LV DAC. When $V_a < 3V$ an error is introduced in the output current because $V_{DS,M2} \neq V_{DS,M1}$. However this error is small because $|V_{DS,M2}| \gg |V_{DS,M1}-V_{DS,M2}|$.

*3.3 Switch Array*

The implemented switch array is shown in Fig. 6(a). The upper switches can be implemented using PMOS HV transistors. The minimum gate length (1.5μm) permitted by the process was chosen to provide low on-resistance. A gate width of 50μm was chosen.

The lower switches are implemented using NMOS HV transistors. The gate lengths and widths were chosen 1.5μm and 15μm, respectively. It should be noted that at node *n* the voltage with respect to ground can become negative because of the charging of $C_L$ and the subsequent reversed current direction during the charge cancellation phase. Therefore a substrate isolated Schottky diode is placed in series to prevent leakage between the substrate and the drain of the transistor. Switch $S_3$ is implemented using back-to-back PMOS transistors with their source terminals biased at $V_{DD}$. The back-to-back configuration is necessary in order to allow for current flow in both directions. Finally, a standard cross-coupled level shifter is used to convert a LV control signal into a HV control signal, $V_{control}$ [19].

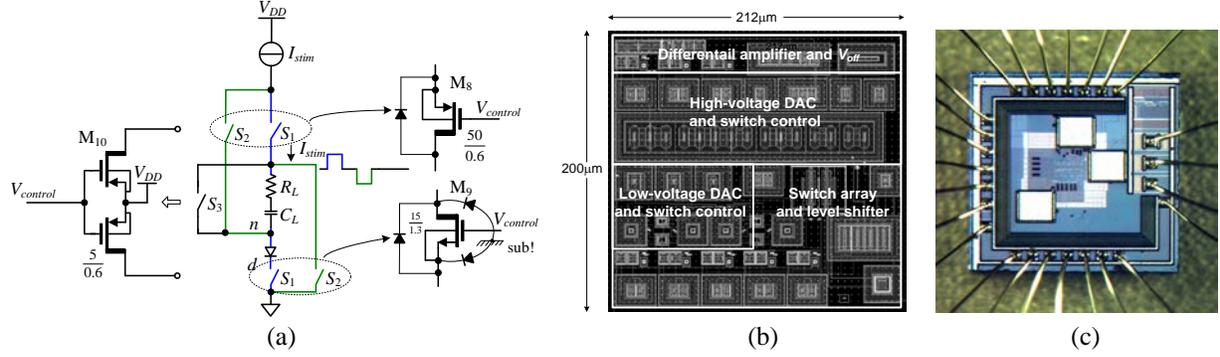

Figure 6: (a) Switch array, (b) layout capture and (c) micrograph.

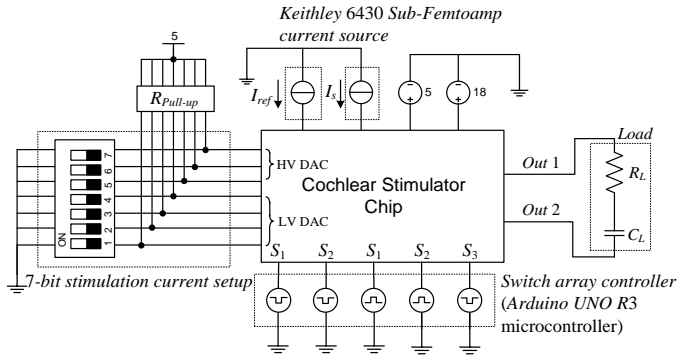

Figure 7: Measurement setup.

Table 1
Transistor dimensions

| MOSFET | $W$ [μm] | $L$ [μm] |
| --- | --- | --- |
| $M_1$, $M_2$(m=1), $M_4$, $M_5$, $M_{11}$, and $M_{12}$ | 5 | 5 |
| $M_6$, $M_7$, and $M_{16}$(n=1) | 0.5 | 0.18 |
| $M_{13}$ | 5 | 0.4 |
| $M_3$, $M_{14}$, and $M_{15}$ | 5 | 0.6 |

The dimensions of the transistors are indicated in Table 1 and were selected to suit their application in fully implantable cochlear implants.

## 4. Measurement results

The stimulator circuit has been implemented in AMS 0.18μm HV CMOS technology. The active area is approximately 200 μm × 210 μm. The layout capture and the micrograph of the chip are depicted in Fig. 6(b) and Fig. 6(c), respectively. From the layout it can be seen that the HV transistors dominate the area. The area was minimized by implementing multiple HV transistors in the same deep n-well whenever possible, e.g. in the HV DAC.

The measurement setup is presented in Fig. 7. An external 18V supply is used for the main circuit. A 5V supply is used for powering the pad ring of the chip. The control of the HV and LV DACs were set by DIP switches. $I_{ref}$ and $I_s$ were set at 10μA by a Keithley 6430 Sub-Femtoamp source meter. Finally, 5 digital outputs from an Arduino UNO microcontroller board were used in order to control the timing and current direction of the switch array.

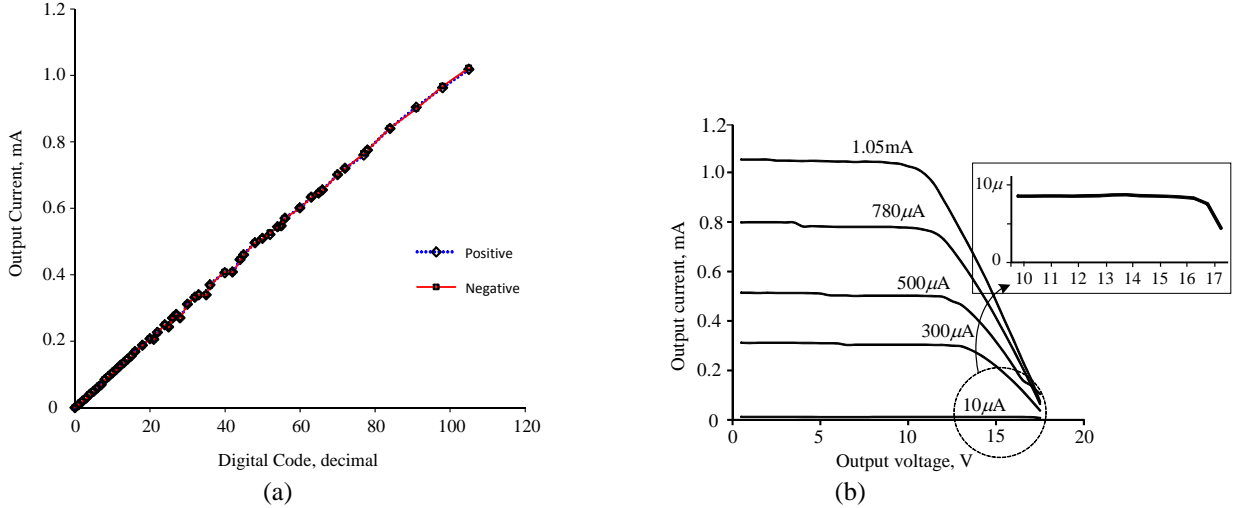

Figure 8: (a) Output current of the positive and negative stimulation direction. (b) Measured output characteristic of the stimulator.

### 4.1 Output DC Characteristic

First the accuracy of the stimulation current was measured. For this, the load was chosen to be a single resistor $R_L$=10 kΩ without a capacitor. The output current was measured using a Keithley 6430 Sub-Femtoamp source meter. Fig. 8(a) shows the measured output current versus full-scale digital input code for the positive and negative stimulation direction, respectively. The current values in both positive and negative directions are almost identical, which is required for proper charge balancing. The percentage of current mismatch is 0.02% and 0.3% at the minimum and maximum output current, respectively. The average percentage of current mismatch over the entire digital input code is 0.05%. This is mainly because of the mismatch error at the most significant bit transistors.

The measured output current versus output voltage at several stimulation current levels is shown in Fig. 8(b). The output voltage was varied from 0 to 18V with a 0.5V step size by using a Keithley 6430 Sub-Femtoamp source meter as a voltage source. The voltage headroom is about 12.5V and 16.5V at maximum and minimum stimulation current, respectively. This value is mainly limited due to the relatively small dimensions of the HV DAC transistors. The voltage headroom can be increased by increasing the width of the HV DAC transistors. Moreover, the voltage headroom also depends on the voltage across the switch array. This voltage is related to the dimensions of the HV switch transistors in order to allow the maximum stimulation current to pass through. In this design it also depends on the voltage across the Schottky diode. This diode has a small (0.4V) voltage drop.

The output resistance calculated from the measurement results plotted in Fig. 8(b) is about 33 MΩ and 500kΩ at $I_{stim}$=10µA and 1.05mA, respectively.

### 4.2 Biphasic Stimulation

Next, the chip was measured while providing a programmable biphasic stimulation current in the range of $I_{stim}$=10µA (minimum) to 1.05mA (maximum). The digital outputs from the microcontroller board are used to control the timing of the biphasic waveform. The pulse widths of $t_a$, $t_i$ and $t_c$ were set to 50µs, and the total cycle time to 600µs. The measured output voltage for $I_{stim}$ = 50µA and 1.05mA at a 10 kΩ resistive load are shown in Fig. 9(a).

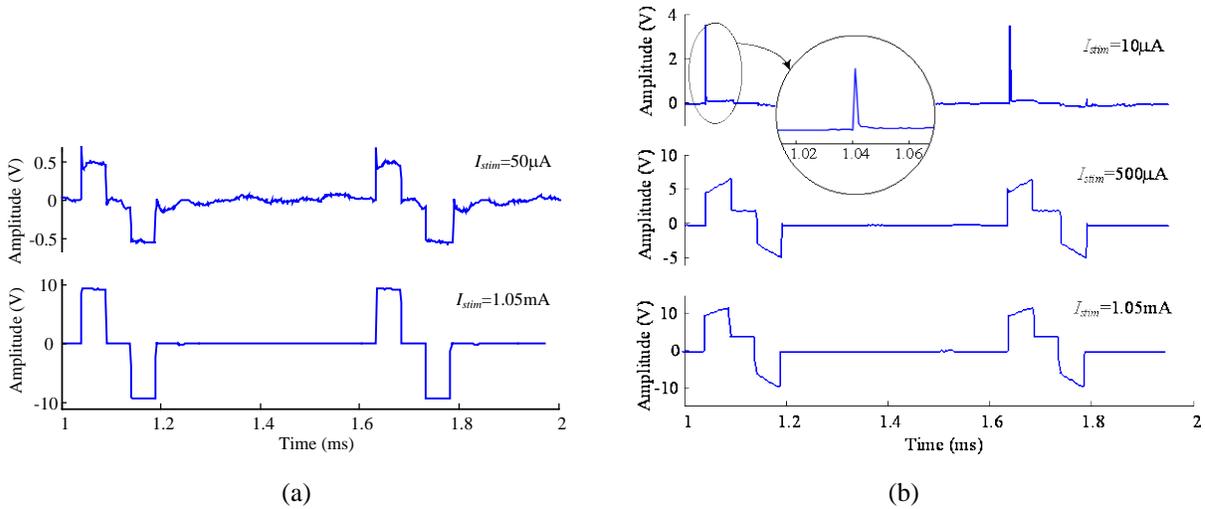

(a)                        (b)

Figure 9: (a) Biphasic output voltage for a 10kΩ load and (b) for a 10kΩ+10nF load.

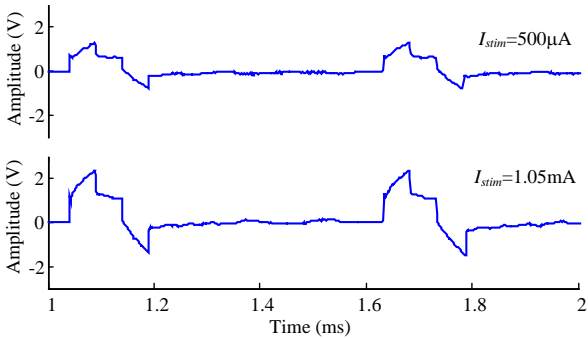

Figure 10: Output voltages for CI electrode load in 0.9% saline solution.

**Table 2**
Charge error and dc current error for $R_L$=10kΩ, $C_L$=10nF

| Stimulation current | Charge error | DC current error |
|---|---|---|
| 10µA (50µs) | 0.50pC | 0.83nA |
| 500µA (50µs) | 0.45pC | 0.75nA |
| 1.05mA (10µs) | 1pC | 1.60nA |

Fig. 9(b) shows the measured biphasic output voltages of $I_{stim}$ across a 10kΩ+10nF load. The spikes due to switching, and consequently settling of the stimulator current sources (see the magnified output voltage in Fig. 9(b) do not contribute to significant charge mismatch, as will be discussed in the next section.

In order to test the stimulator in a realistic situation, the load was changed to a CI electrode array in 0.9% saline solution. Clarion HiFocus Cochlear electrodes were used. These are platinum iridium electrodes, produced by Advanced Bionics, used for studies in animal cochleae and each electrode has an area of approximately 0.2 mm$^2$. Fig. 10 shows the measured output voltage across the two electrode sites at $I_{stim}$=500µA, and 1.05mA. As can be seen from Fig. 10, the circuit works as expected. The output voltage at the end of each stimulation cycle remains constant and goes to zero without creating any voltage accumulation.

### 4.3 Charge Error

The residual voltage ($V_{residual}$) at the end of the stimulation cycle has been measured to determine the remaining charge imbalance. It was measured by connecting an instrumentation amplifier (AD826) in parallel with $C_L$. The output of the amplifier was connected to a 20-bit analog to digital converter (ADC) card (APPLICOS model ATX7006). A shielding enclosure (ground connected Faraday cage) was used to reduce the amount of noise picked up from the environment. A computer running APPLICOS ATXView software was used to acquire, store and analyze the data. The average DC offset voltage and noise in the data acquisition hardware was measured before each residual voltage measurement and subsequently subtracted from the acquired data. The charge error can be calculated by multiplying the measured residual voltage at the end of stimulation cycle with the capacitive load value ($Q_{error} = C_L V_{residual}$). The dc current error is subsequently determined by dividing the charge error by the stimulation cycle time ($I_{dc} = Q_{error}/t_{stim}$). Table 2 shows the computed charge errors and dc current errors for several values of $I_{stim}$ at 10kΩ+10nF load. The pulse width was set to 50μs and the stimulation cycle to 600μs. For $I_{stim}$ =1.05mA the pulse widths $t_c$ and $t_a$ were chosen to be 10μs to prevent clipping of $I_{stim}$. The results show that the charge error and DC current error stay well below the safety limits. The specified industry limit on current mismatch in cochlear implants is 25 nA [20].

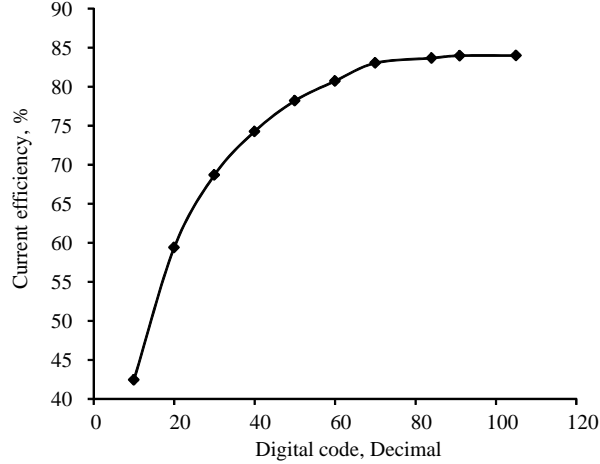

Figure 11: Current efficiency.

### 4.4 Current and Power Efficiency

For the maximum output current ($I_{stim}$=1.05mA) through the maximum load ($R_L$ = 10kΩ), the current efficiency (defined by the ratio of the load current and the supply current) is 87% as shown in Fig. 11. The maximum power efficiency is found to be 61%. The power consumption is dominated by the bias sources in the differential amplifier (30μA) and the current through the DACs, depending on the number of bits enabled in the LV DAC (ranging from 40~158μA). The measured quiescent current is 210μA which is limited by the biasing of $M_3$. Note that all these bias sources can be switched off when stimulation is not active, yielding very low static power consumption.

### 4.5 Multichannel Operation

The proposed stimulator can be used for multichannel stimulation as shown in Fig. 12. The reference circuit provides biasing quantities for the LV DAC, HV DAC and amplifier $Z_m$, which can be shared among stimulation channels. The control logic for setting the 7 bit current amplitude and 3 bit current direction/pulse width are received from a programmable controller.

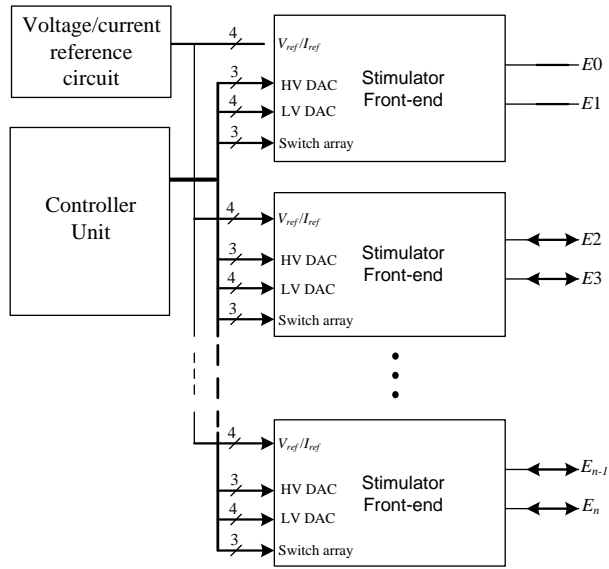

Figure 12: Multichannel stimulator.

## 5. CONCLUSION

A compact programmable biphasic stimulator chip for cochlear implants has been presented in this paper. A double loop negative feedback topology was employed to increase the output impedance. The circuit can deliver stimulation amplitudes in the range of 10μA~1.05mA for a wide range of electrode-tissue impedances: $R_L$=1kΩ~10kΩ, $C_L$=1nF~10nF. The current error (<1.6nA) was found to be well below the safety limits. It consumes a very small chip area (0.042mm$^2$) enabling many stimulation channels on a single die.

**Acknowledgment**

The authors would like to thank Wil Straver and Ali Kaichouhi for technical support during the chip measurement.

**References**

[1] B.S. Wilson, " Engineering design of cochlear implants," in *Auditory prostheses: Cochlear implants and beyond*. Zeng FG, Popper AN, Fay RR, Ed. New York (NY): Springer; 2004. p. 14–52.

[2] J. C. Lilly, J. R. Hughes, E. C. Alvord, Jr., and T. W. Galkin, "Brief, noninjurious electric waveforms for stimulation of the brain," *Science*, vol. 121, pp. 468–469, 1955.

[3] S. C. DeMarco, W. Liu, P. R. Singh, G. Lazzi, M. S. Humayun, and J. D. Weiland, "An arbitrary waveform stimulus circuit for visual prosthesis using a low-area multibias DAC," *IEEE J. Solid-State Circuits*, vol. 38, no. 10, pp. 1679–1690, Oct. 2003.

[4] M. Sivaprakasem, W. Lui, M. S. Humayun, and D. J. Weiland, 'A variable range bi-phasic current stimulus driver circuitry for an implantable retinal prosthesis device," *IEEE J. Solid-State Circuits*, vol. 40, no. 3, pp 763-771, Mar. 2005.

[5] E. K. F. Lee and A. Lam, "A matching technique for biphasic stimulation pulse," in *Proc. IEEE ISCAS*, 2007, pp. 817–820.

[6] M. Ortmanns, N. Unger, A. Rocke, M. Gehrke, and H. J. Tietdke, "A 0.1 mm$^2$, digitally programmable nerve stimulation pad cell with high voltage capability for a retinal implant," in *Proc. IEEE ISSCC*, 2006, pp. 89–98.

[7] M. Ghovanloo and K. Najafi, "A compact large voltage compliance high output impedance programmable current source for implantable microstimulators," *IEEE Tran. Biomed. Eng.*, vol. 52, pp. 97-105, Jan. 2005.

[8] J. J. Sit, and R. Sarpeshkar, "A low-power blocking-capacitor-free charge-balanced electrode-stimulator chip with less than 6nA DC error for 1-mA full-scale stimulator," *IEEE Tran. Biomedical Circuits Syst.*, vol. 1, no. 3, pp.172-183, Sep. 2007.

[9] M. Ghorbel, M. Samet, A. Ben Hamida, J. Tomas, "A 16-electrode Fully Integrated and Versatile CMOS Microstimulator Dedicated to Cochlear Implant", *Journal of Applied Sciences*, vol. 6, no.15, pp. 2978–2990, Sep. 2006.

[10] L. Zeng and X. Yi and S. Lu and Y. Lou and J. Jiang and H. Qu and N. Lan and G. Wang, "Design of a high voltage stimulator chip for a stroke rehabilitation system." *Engineering in Medicine and Biology Society (EMBC), 2013 35th Annual International Conference of the IEEE*. IEEE, 2013.

[11] X. Liu, A. Demosthenous, and N. Donaldson, "An integrated implantable stimulator that is fail-safe without off-chip blocking-capacitors," *IEEE Trans. Biomed. Circuits Syst.*, vol. 2, no. 3, pp.231–244, Sep. 2008.

[12] H. Chun, Y. Yang, and T. Lehmann, "Safety ensuring retinal prosthesis with precise charge balance and low power consumption," *IEEE Trans. Biomed. Circuits Syst.*, vol. 8, no. 1, pp.108-118, Feb. 2014.

[13] E.T. McAdams and J. Jossinet, "Nonlinear transient response of electrode-electrolyte interfaces", *Medical & Biological Engineering & Computing*, Vol. 38, pp 427-432, 2000.

[14] K. E. Jones and R. A. Normann, "An advanced demultiplexing system for physiological stimulation," *IEEE Trans. Biomed. Eng.*, vol. 44, no.12, pp. 1210–1220, Dec. 1997.

[15] E. Noorsal, K. Sooksood, Hongcheng Xu, and R. Hornig, J. Becker and M. Ortmanns, "A neural stimulator frontend with high-voltage compliance and programmable pulse shape for epiretinal implants," *IEEE J. Solid-State Circuits*, vol. 47, no. 1, pp. 244–256, Jan. 2012.

[16] C. Sawigun, W. Ngamkham and W. Serdijn, "A least-voltage drop high output resistance current Source for Neural Stimulation," in *IEEE BioCAS*, Paphos, Cyprus, Nov. 3-5, 2010.

[17] W. Ngamkham, C. Sawigun, and W. A. Serdijn, "Biphasic Stimulator Circuit for a Wide Range of Electrode-Tissue Impedance Dedicated to Cochlear Implants," in *Proc. IEEE International Symposium on Circuits and Systems*, Seoul, Korea, May 20 - 23, 2012.

[18] N. S. Lawand, W. Ngamkham, G. Nazarian, P. J. French, W. A. Serdijn, G. N. Gaydadjiev, "An improved system approach towards future cochlear implants, *Proc. IEEE EMBS*, Osaka, Japan, July 2013.

[19] Y. Chin Ping and C. Yu Chien, "A Voltage Level Converter Circuit Design with Low Power Consumption" *IEEE, The 6th IEEE International Conference on ASIC*, October 2005.

[20] C. Q. Huang, R. K. Shepherd, P. M. Carter, P. M. Seligman, and B. Tabor, "Electrical stimulation of the auditory nerve: Direct current measurement in vivo," *IEEE Trans. Biomed. Eng.*, vol. 46, pp.461–470, Apr. 1999.